\documentstyle[12pt]{article}
\newcommand{\be}{\begin{equation}}
\newcommand{\ee}{\end{equation}}
\newcommand{\by}{\begin{eqnarray}} 
\newcommand{\ey}{\end{eqnarray}}
\newcommand{\om}{\omega}
\newcommand{\de}{\delta}
\newcommand{\dr}{\left ({\delta \rho\over {\rho}} \right )}
\newcommand{\tin}{\int {d^3 k\over {(2\pi)^3}}}
\newcommand{\lb}{\left ( }
\newcommand{\rb}{\right ) }
\newcommand{\lc}{\left \langle }
\newcommand{\rc}{\right \rangle }

\newcommand{\bm}{\bibitem}
\begin{document}
\rightline{SINP-TNP/96-20}
\vspace{1cm}
\centerline{\large \bf Density perturbation in extended inflation}
\vspace{1.5cm}
\centerline{S. Mallik $^a$ and D. Rai Chaudhuri $^b$ }
\centerline{\it 
$^a$ Saha Institute of Nuclear Physics, 1/AF, Bidhannagar, Calcutta 700064,
India}
\centerline{\it
$^b$ Acharya Jagadish Chandra Bose College, 1/1B, AJC Bose Road,
 Calcutta 700020, India}

\vspace{1cm}
\begin{abstract}
We examine the calculation of density perturbation at large scales in the
inflationary scenario. The formula for its magnitude is reviewed from first
principles and applied to the original model of extended inflation. Our
estimate is an order of magnitude bigger than the earlier ones due to the
difference in the time at which the primordial fluctuation is evaluated and
to the inclusion of terms neglected earlier in the standard formula for
density perturbation.
\end{abstract}
\section{Introduction}
\setcounter{equation}{0}
      
One of the remarkable features of inflationary models
\cite{Blau} is that, besides
solving the classical problems of large scale homogeneity and flatness, it
also addresses the problem of creating the seeds of density inhomogeneity in
the early universe. These are the quantum fluctuations in energy density
during the inflationary epoch, which might have evolved to produce
the observed density inhomogeneity in the universe. Much work has naturally
been devoted to a study of the evolution of the initial density perturbation
\cite{Hawk, Gupi, Bardeen}.

The difficulty here is that of constructing a realistic model \cite{Kolb}.
In the original, the `old' inflationary model
\cite{Guth}, the bubbles of
true phase, nucleated during the strongly first order phase transition,
cannot percolate. The slow-rollover models 
\cite{Linde} require too small a coupling 
constant for reproducing the observed density fluctuation to cause problem
of reheating. Finally the so-called `chaotic' models 
\cite{Chaotic} require initial
conditions having no resemblance to the thermal equilibrium condition
(or definite departures therefrom), usually believed to prevail in the
early universe.

The extended inflationary models \cite{La}
are similar to the `old' model with the Einstein gravity replaced by the
theory of Jordan \cite{Jordan} and
of Brans and Dicke \cite{Brans}.  The idea is to take
advantage of the time dependence of 
gravitational `constant' to solve the bubble
nucleation problem. Though a realistic model still eludes 
construction \cite{Liddle}, 
further work in this direction  is expected to lead to such
a model without violating astrophysical observations \cite{Reas}.

In this work we examine the evaluation of density perturbation in the
inflationary epoch
\cite{Bunch}. We first relate the density perturbation to the quantum
fluctuations in the energy density operator from first principles
\cite{Kotu, Padma} . We then 
consider the original model of extended inflation in the so-called Einstein
frame \cite{Ho}. By restricting the energy density 
operator to terms linear in the
scalar field, the quantum fluctuations in the density reduce essentially to
the two point function of the scalar field. The density perturbation turns
out to be an order of magnitude bigger than the earlier evaluations
\cite{Kosatu, Guja}.

In sec. 2 we derive the formula for the density perturbation from
first principles. 
In sec. 3 we consider the original model of extended inflation and
derive the two point function for the scalar field. This is used to evaluate
the density perturbation in sec. 4. In sec. 5 we identify the points of
difference between the present and the earlier evaluations leading to an
order of magnitude difference. Finally in sec. 6 we summarise our work
emphasising these differences.

\section{Density fluctuation formula}
\setcounter{equation}{0}
We begin by reviewing the definitions of density inhomogeneity and quantum
fluctuation to relate them from first principles \cite{Kotu, Padma}. 
Let $\rho (\vec{x},t)$ be the energy density field of the Universe, which we
assume for simplicity to be confined within a (large) volume $V$. The mean
square fluctuation in $\rho (\vec{x},t)$ is defined as
\be
{\dr}^2 = \lc \lb {{\rho (\vec{x},t) -\bar{\rho} (t)}\over
 {\bar{\rho} (t)}}\rb ^2\rc _{\vec{x}},
\ee
where ${\langle \cdots \rangle}_x$ denotes average over all space and
$\bar{\rho} (t) = \langle \rho (\vec{x},t)\rangle _x$, the averaged, homogeneous
background density. One writes
\be
\rho (\vec{x},t) = \bar{\rho} (t)(1+\de (\vec{x},t)),
\ee
where the density contrast $\de (\vec{x},t)$ has the Fourier expansion
\be
\de (\vec{x},t) ={1\over \sqrt{V}}\sum_{k}  \de_k (t) 
e^{-i\vec{k} \cdot \vec{x}}.
\ee
Then the mean square fluctuation (2.1) becomes
\be
\dr^2 = {1\over V} \sum_{k} |\de_k (t)|^2
\rightarrow \int{{k^3 |\de_k (t)|^2}\over {2\pi^2}} d (\ln k).
\ee
The so-called fluctuation power per logarithmic interval
in wavenumber is defined as
\be
\dr^2_k = {k^3|\de_k (t)|^2\over {2\pi^2}}.
\ee

For our purpose it is useful to consider a related quantity, 
namely, the mean square
mass fluctuation on a given length scale. It measures the mass fluctuation
within a certain volume $v$ by averaging 
the squared excess mass in it over all points $x_0$
throughout the volume $V$ of the universe. To avoid sharp boundary, one
smears $v$ with a Gaussian window function. 
The mass within such a smeared sphere placed at $\vec{x}_0$ is
\be
m_l (\vec{x}_0 ,t) = \int d^3 y e^{-y^2 /2l^2}  \rho (\vec{x}_0 +\vec{y},t).
\ee
The mean square mass fluctuation on length scale $l$ is then
\be
\left ( {\de m\over m} \right )^2_{l,c} = {\lc \lb 
{{m_l (\vec{x}_0 ,t) - \bar{m}_l (t)}
\over {\bar{m}_l} (t)} \rb^2 \rc}_{\vec{x}_0} .
\ee
Here  $\bar{m_l} (t)= <m_l (\vec{x_0},t)>_{\vec{x_0}}$,
the average mass
obtained by replacing $\rho$ by $\bar{\rho}$ in (2.6). 
The subscript c stands for classical and indicates the phenomenological
nature of the evaluation. Inserting (2.3) in (2.6) we get 
\be 
\left ({\de m\over m} \right )^2_{l,c} = \tin  |\de_k (t)|^2 e^{-k^2 l^2}.
\ee

Now suppose that this density inhomogeneity has a quantum origin
in the inflationary epoch. Then
it should be calculable by considering the quantum fluctuations  
in the energy density operator, $T_{00}$. This operator will be written
explicitly for the specific Lagrangian of extended inflation considered in
the next section. Here we write it as a sum of two parts,
\be
T_{00} (\vec{x},t) =\bar{\rho} (t) + U(\vec{x},t),
\ee
where $\bar{\rho}$, the classical part, is to be identified with the
background density considered above and $U$, the quantum part, depends on
the quantum field operator.
 
In analogy with (2.6) let us define the mass operator,  
\be
\hat{m}_l (t) = \int d^3 x e^{-x^2 /2 l^2} T_{00} (\vec{x},t),
\ee
where the hat on $m$ emphasises that it is an operator. 
The mean squared fluctuation in mass in the smeared sphere is
\be
{\left ( {\de m\over m} \right )}^2_{l,q} = \lc \lb {{\hat{m}_l (t)
-\bar{m_l} (t)}\over
\bar{m_l} (t)} \rb ^2 \rc,
\ee
where $\langle \cdots \rangle $ denotes the quantum mechanical expectation
value in the state with energy density $\bar{\rho}(t)$, which corresponds to
the vacuum of the conventional quantum field theory. The subscript $q$
distinguishes it from the phenomenological evaluation in (2.7-8).
Inserting the Fourier transform of the two point function of $U$
\be
\langle U(\vec{x},t) U(\vec{x'},t) \rangle 
= \tin e^{-i\vec{k} \cdot (\vec{x}-\vec{x'})} D_k (t),
\ee 
in eqn (2.11), we get
\be
\left ( {\de m\over m} \right )^2_{l,q} = \tin e^{-k^2 l^2} {D_k (t)\over 
{\bar{\rho}^2 (t)}}.
\ee

If we now identify the spatial averaging in (2.7) with the quantum mechanical
expectation value in (2.11), we get the desired relation,
\be
|\de _k (t)|^2 = {D_k (t)\over {{\bar{\rho}}^2 (t)}},
\ee
allowing us to calculate the density fluctuation during inflation,
when it is still within the Hubble length (causal horizon).
Its evolution as superhorizon-sized perturbation, 
from the time $t_h$ when it leaves the Hubble radius until its re-entry
within it later, is described by the constancy of the quantity
$ \zeta ={\de}_k (t) /(1+ \bar{p} / \bar{\rho})$ \cite{Bardeen}.
 Assuming radiation dominance at
the time of re-entry, we finally get the density perturbation at that time
as
\be
\left ( {\de \rho \over \rho} \right)_H = \frac{4 \sqrt{ k^3 D_k (t_h)}}
{3 \sqrt{2} \pi 
(\bar{\rho}+\bar{p})_{t_h}}. 
\ee

\section{Extended inflation in Einstein frame}
\setcounter{equation}{0}

Because of the nonminimal coupling of the
scalar field in the original 
Brans-Dicke (BD) Action \cite{Brans}, it is not possible to
do calculations with it in the canonical framework of quantum field theory.
It 
has, however, been shown 
\cite{Ho} that an appropriate Weyl rescaling \cite{Birrell} can
transform this action to a form where both the gravity and the kinetic term
in the scaled BD field are canonical. In this so-called
Einstein frame, the rescaled action becomes 
\be
S=\int d^4 x \sqrt{g} \left ({R\over{16\pi G}} + {1\over 2}g^{\mu \nu}
\partial_{\mu}\Psi \partial_{\nu}\Psi -V (\Psi) \right)
\ee
where $\Psi (x)$ is the BD field in the Einstein frame and
\[ V(\Psi) =M^4 e^{-2\Psi /\psi_0}, \qquad \psi_0 =\sqrt{{2\om +3}\over 16\pi}
m_P ,\]
$ m_P \equiv G^{-1/2} $ being the present value of the Planck mass.
This form for the potential function results from assuming the matter
Lagrangian in the original 
(Jordan) frame to be dominated by the false vacuum energy
density, $M^4$. The dimensionless parameter $\omega$ appears in the original
BD Lagrangian.

We decompose the BD field $\Psi$ into a homogeneous, classical field $\psi$
and a quantum field $\phi$,
\be
 \Psi (\vec{x},t) =\psi (t) +\phi (\vec{x},t). 
\ee
In the spatially flat FRW metric, $ ds^2 =dt^2 -a(t)^2\vec{ dx}^2$,
the classical equations of motion for $\psi (t)$ and the scale factor
$a(t)$ are
\be
\ddot{\psi}+ 3 {\dot{a}\over a} + {d V\over{d\psi}} =0 
\ee
\be
\left (  {\dot{a}\over a} \right )^2 = {8\pi \over{3 m_P^2}} \left (
{1\over 2} \dot{\psi} ^2 + V(\psi) \right)
\ee
In the Einstein frame, the extended inflation resembles  a slow-rollover 
inflation off an exponential potential \cite{Kosatu}.

The solution to the coupled equations are \cite{La, Ho}
\be
\psi (t) =\psi(0) +\psi_0 \ln (1+Ct), 
\ee
\be
a(t) =a(0) (1 +Ct)^p, \qquad p=(2\omega +3)/4.
\ee
where
\[C= {2M^2\over q\omega m_P} e^{-\psi(0)/ \psi_0}, \qquad q\om =
\sqrt{(6\om +5)(2\om +3)\over {32\pi}} .\]
We note that the Hubble length in this model is,
\be
H(t)^{-1} =\lb {\dot{a}\over a} \rb ^{-1} ={{1+Ct}\over Cp}.
\ee

We now turn to the quantum theory. Expanding the Lagrangian around
the classical field $\psi$, one gets an (infinite) series in powers of the
quantum field $\phi$, though the coefficient of expansion $(M / m_P)$
is small. Such a Lagrangian is, of course, not perturbatively
renormalisable. Here we simply reject all higher order
terms retaining, in fact, only the lowest (second) order terms to get
\be
S_q =\int d^3x dt a^3(t) \left ( {1\over 2} \dot{\phi}^2 -{1\over {2 a^2}}
(\nabla \phi)^2 -{1\over 2} {\mu}^2 (t) \right ),
\ee
where
\[ {\mu}^2 (t) ={d^2 V\over{d {\psi}^2}} = {2(3p-1)\over p^2} H^2(t).\]
 
We calculate the Feynman propagator for $\phi$ in the classical background
field $\psi (t)$. Its Fourier transform in 3-space, 
\be
\langle T \phi (\vec{x},t) \phi (\vec{x'},t') \rangle =\tin 
e^{-i \vec{k} \cdot (\vec{x}-\vec{x'})} G_k (t,t'),
\ee
satisfies the inhomogeneous equation
\be
\left ( {d^2\over {d t^2}} +3 {\dot{a}\over a} {d\over {dt}}
+ {k^2\over a^2} + {\mu}^2 (t)
\right ) G_k (t,t') = -{i\over {a^3 (t)}} \delta (t-t').
\ee
The Green's function can be constructed by the familiar procedure. Let
$f^{(+)} (t), f^{(-)} (t)$ be two linearly independent mode 
functions satisfying the
homogeneous part of the above equation and incorporating the boundary
conditions that $ f^{(\pm)}$ contain respectively positive and negative 
frequencies during the initial period for not too low physical momenta
\cite{Foot1}. Their normalisation is fixed by 
the value of the Wronskian of the two solutions derivable from (3.10).
Then  we have 
\be
G_k (t,t') =f^{(+)} (t) f^{(-)} (t') \theta (t-t') +f^{(-)} (t) f^{(+)} (t')
 \theta (t'-t).
\ee
The equation for the mode functions can be solved exactly. With a change of
variable \cite{Abbott}, $\tau =(1+Ct)^{1-p}/(p-1)$, it becomes
\be
\left ({d^2\over {d\tau^2}} -{2p\over {p-1}} {1\over \tau}{d\over {d\tau}}
+{\kappa}^2 + {2(3p-1)\over {(p-1)^2 \tau^2}} \right ) f^{(\pm)} (\tau) =0,
\qquad \kappa ={k\over {a(o) C}}.
\ee
It is now easy to cast it in the standard form of Bessel's equation. We get 
\be 
f^{(+,-)} (t) = {1\over 2} \sqrt{{p\over {p-1}} { \pi\over H}}
{H_{\nu}^{(1,2)} (z)\over {a(t)^{3/2}}},
\ee
where $H_{\nu}^{(1,2)} (z)$ are the Hankel functions of the first and second
kind respectively \cite{Abram, Foot2} and
\[ \nu = \frac{\sqrt{3(3p-1)(p-3)}}{2(p-1)}, \qquad
 z=\kappa \tau ={p\over {p-1}}{k\over {a H}}.  \]
Two point functions involving  $\phi$ and $\dot {\phi}$ are immediately
obtainable from (3.11) and (3.13).

We also write here the expressions for the energy density and pressure,
which can be obtained directly from the energy-momentum tensor,
\be
T_{\mu \nu} = \partial_{\mu} \Psi \partial_{\nu} \Psi -g_{\mu \nu} 
({1\over 2} g^{\alpha \beta} \partial_{\alpha} \Psi \partial_{\beta} \Psi
- V(\Psi)) ,
\ee
for the action (3.1). We have already split the energy density as the sum of
a homogeneous, classical part and a quantum part in (2.9). Inserting (3.2)
in (3.14), they are
\be
 \bar{\rho} ={1\over 2} \dot{\psi}^2 + M^4 e^{-2\psi (0) /{\psi_0}} 
= {3\over 8 \pi} m_p^2 H^2,
\ee
\by
U &=& -{M^4\over {\psi_0^2}} \phi +  \dot{\psi} \dot{\phi} \nonumber \\
  &=& -{{m_p H^2}\over {\sqrt{{\pi} (2\om +3)^3}}} [(6\om +5) \phi + (2\om-1)
\tau {d\over {d\tau}} \phi].
\ey
In $U$ we have only retained terms linear in $\phi$. Higher order terms like 
$\phi ^2$ would give rise to loops and hence divergence in the expression
for the density fluctuation. As already stated, such terms are multiplied by
small coefficients. Thus in any reasonable renormalisation scheme , such
terms are expected to produce small contributions.
We also note the expression for the homogeneous pressure,
\be
\bar{p} = {1\over 2} \dot{\psi}^2 -M^4 e^{-2\psi(0)/{\psi_0}}
=-{1\over {8\pi}} {{6\om +1}\over {2\om +3}} m_P^2 H^2.
\ee
\section{Evaluation of $\delta \rho / {\rho}$}
\setcounter{equation}{0}
Since the energy density operator $U(\phi)$ is linear in $\phi$, it is
simple to express the two point function of $U(\phi)$ in terms of the mode
functions $f^{(\pm)} (t)$. Noting the symmetry of Hankel functions,
$H_{\nu}^{(2)} (z) = H_{\nu}^{(1)*} (z) $ for real $z$, we can write its
Fourier transform (2.12) as
\by
D_k (t) &=& {m_p^2 H^4\over {\pi (2\om +3)^3}} |(6\om +5) f^{(+)} (\tau)
+(2\om -1) \tau {d\over {d\tau}}f^{(+)}|^2, \nonumber  \\
        &=& {m_p^2\over {4(2\om -1) (2\om +3)^2}} \lb {H(t)\over a(t)}\rb ^3
 |F(t)|^2,
\ey
where 
\[ F(t)={3\over 2} (6\om +5) H_{\nu}^{(1)} (z) + (2\om -1) z {d\over {dz}}
H_{\nu}^{(1)} (z) .\]

Let $\lambda _{phys} (t)$ be the physical wavelength characterising the 
density perturbation at time $t$,
belonging to the comoving wavenumber $k$, $\lambda_{phys} (t) =2\pi a(t)/k$. 
 At time $t_h$
during inflation, when it equals the Hubble length, its magnitude is 
$ \lambda _{phys} (t_h) =H^{-1} (t_h)$, so that $k/(aH)|_{t_h} =2\pi.$
Then eq.(2.15) becomes
\be
{\dr}_H ={4 {\pi}^{3/2} \over{3\sqrt{2\om-1}}} {H(t_h)\over m_p} |F(t_h)|.
\ee
  
At time $t_h$ the argument of the Hankel function becomes
 $z(t_h)=2\pi p/(p-1)$, a value large enough to justify the use of 
asymptotic expansion
\cite{Abram}. Retaining terms in $z$ upto the first nonleading
one, it is simple to evaluate $F(t_h)$ giving
\be 
F(t_h)= 2\sqrt{(2\om-1)(2\om+3)} K(\om), 
\ee
where $K$ is of order unity,
\[K^2(\om) =1+{3(20 \om^2 +48\om +27)\over{4\pi^2 (2\om +3)^2}}. \]

To determine $t_h$, we follow earlier authors \cite{Kosatu, Guja} to make 
 certain simplifying assumptions.  It is assumed that the extended
 inflation ends at time $t_e$ and instantly gives rise to the radiation
dominated era with an initial temprerature $T \simeq M$. Further, since the
gravitational `constant' ceases to vary in the radiation era, the
(classical) BD field $\psi$ is set equal to zero from the time $t_e$
onwards \cite{Foot3}. We then get from (3.5),
\be
{M\over H(t_e)}={q\om\over 2p} {m_p\over M}.
\ee
 The corresponding wavelengths at time $t_h$ and at the present time $t_p$
can be related as
\[ \lambda_{phys} (t_h) ={a(t_h)\over a(t_e)} {a(t_e)\over a(t_p)}
\lambda_{phys} (t_p). \]
The scale factors can now be evaluated to give
\be
{M\over H(t_h)} = \left ( {M/H(t_h)\over{M/H(t_e)}} \right ) ^p T_p
{\lambda}_{phys} (t_p),
\ee
where $T_p= 2.7 K$, the present temperature of the background radiation. 
Solving for $M(t_h)$ and using (4.4) we get
\be
H(t_h)=M \left ( {2p\over q\om} {M\over m_P} \right )^{p\over (p-1)}
\left (T_p \lambda_{phys} (t_p) \right)^{1\over (p-1)}.
\ee
Inserting this value in (4.2) we finally get
\be
{\dr}_H = {8\pi \over 3} K \sqrt{2\pi} [(2\om +3)/2]^{6\om +5 \over
{2(2\om-1)}} (q\om)^{-{{2\om+3}\over {2\om -1}}} (M/m_P)^{2(2\om +1)\over
{2\om-1}} (T_p {\lambda} (t_p))^{4\over {2\om -1}}.
\ee
This result differs from what one finds in the literature \cite{Kosatu,
Guja}. In particular it is larger than the result by
Guth and Jain \cite{Guja} by a factor of 
$8\pi K/3 \simeq 10$. The sources of this difference are discussed in the 
next section.

\section{Comparison with earlier works}
\setcounter{equation}{0}

There are two major differences with earlier works, which contribute to the
enhancement of our result. One is the time at which the mode functions are
evaluated. As already stated, the argument of the Hankel functions around
the time of Hubble length crossing is sufficiently large for the asymptotic
expansion to apply,
\be
H_{\nu} ^{1,2} (z) \rightarrow \sqrt{2\over {\pi z}} e^{\pm iz},
\quad (z \; large).
\ee
Then the fluctuation in these oscillatory Fourier components of the quantum
field is given by
\be
{k^3 |f^{\pm)} (t) |^2\over 2{\pi}^2} = \lb {k\over aH} {H\over {2\pi}}\rb
^2,
\ee
which reduces to just $H^2$ at the time of horizon exit, independently of
the value of $\nu$.

But in the literature, the evaluation at the Hubble length crossing is
{\it{actually}} carried out when the wavelength of the perturbation is much
larger than the Hubble length and the corresponding mode ceases to
oscillate. In this `frozen' state, the argument of the Hankel functions is
small enough for expansion around the origin to apply,
\be
H_{\nu}^{(1,2)} (z) \rightarrow  \pm i{\Gamma (\nu)\over \pi} 
({z\over 2})^{-\nu}, \quad (z \; small).
\ee
Further $\nu$ is assumed very close to $3\over 2$ . Under these conditions the
fluctuation in the quantum field is given by
\be
\left ( {k^3 |f^{(\pm)} (t)|^2 \over {2 \pi^2}} \right )^{(o)}
= \lb{ {p-1}\over p}{H\over {2\pi}} \rb ^2 .
\ee
The superscript $(o) $
indicates evaluation following literature. Thus our expression (5.2), though
not of this form in general, does reduce to this form at the time of Hubble
length crossing but is larger by a factor of $(2\pi p/(p-1))^2$.

The other difference is the omission of the $\dot{\phi}$ term in the energy
density operator. Thus in the literature the energy density fluctuation is
related to the field fluctuation by
\be
\left ({k^3 D_k (t)\over {2\pi^2}} \right )^{(o)} = \left ( {\partial V\over
{\partial \psi}} \right ) ^2
\left ( {k^3 |f^{(\pm)} (t)|^2 \over {2 \pi^2}} \right )^{(o)} . 
\ee
Comparing with (4.1) we see that the neglected term is of the same order of
magnitude as the one retained for $z=2\pi p /(p-1)$.

At this point it is simple to obtain the standard expression for the density
fluctuation and verify the enhancement factor. For the slow-rollover
scenario, $(\partial V/\partial \psi)$ may be estimated from the classical
equation of motion (3.3) as $(\partial V/\partial \psi) = -3H\dot{\psi}$.
Then inserting (5.4-5) into (2.15) we recover the standard formula,
\be
{\dr}_H^{(o)}= \lb {H^2\over {\dot{\psi}}} \rb _{t_h},
\ee
on ignoring a factor of $2(p-1)/{\pi p}$. It can be immediately evaluated to
give
\be
{\dr}_H^{(o)} = \sqrt{\pi (2\om +3)} {H\over m_P}
\ee
 This standard result is clearly smaller than our result (4.2-3) by the
factor $8\pi K/3 $.
\section{Discussion}

We review from first principles the formula for the density perturbation due
to quantum fluctuations in the energy operator. We then evaluate the
resulting formula in the context of the original model of inflation in the
Einstein frame, using the simplifying assumptions made by the earlier
authors.

The present evaluation of the density fluctuation differs from the earlier
ones in two respects.  One is the time at which the perturbation in the
inflationary epoch is evaluated. We evaluate it exactly at the time when
the characteristic wavelength starts growing bigger than the Hubble
length. Then the mode functions are still oscillatory and physical
quantities admit conventional interpretation. In the literature, however, 
the perturbation is actually evaluated when the wavelength has grown several
times bigger than the the Hubble length and the mode functions have ceased
to oscillate.
 
Although evaluated at two different times with qualitatively different mode
functions, the density fluctuations turn out to have the same, nearly scale
independent, spectum. It is interesting to note that our evaluation leaves
the index $\nu$ of the Hankel function free, while that in the literature
requires $\nu$ to be equal to ${3\over 2}$ to arrive at this spectrum. But the two
evaluations do make a difference in the magnitude of the fluctuation, our
result being greater by a factor of $2\pi$.
 
The other point of difference with the earlier evaluation is the inclusion
of the $\dot{\phi}$ term in $T_{00}$. In our way of evaluation, this term is
as important as the $\phi$ term. These two sources of difference go to make
our result larger by about a factor of 10.

We hope to have made it clear that these sources of enhancement 
in our calculation over the earlier ones,
although discussed here in the context of a specific model, is,
in fact, quite general. But to arrive at the physically correct magnitude of
$(\delta \rho /{\rho})_H$, we need a better knowledge of the evolution of
the density perturbation than what the approximate constancy of $\zeta$
suggests. Pending this investigation, a comparison of our result with the
earlier ones indicates the magnitude of uncertainty in such a calculation.

\bigskip

One of us (SM) would like to thank Professor Paul J. Steinhardt for earlier 
discussions on related topics. He also gratefully acknowledges the hospitality
at the Institute for theoretical Physics, University of Bern, which made
these discussions possible.

\end{document}